\documentclass[12pt,draftcls,onecolumn]{IEEEtran}

\usepackage{latexsym,url}
\usepackage{amsmath,amssymb,array,graphicx,verbatim}
\usepackage{setspace}
\usepackage{amsfonts}
\usepackage{amsmath,amssymb}
\usepackage{cite}
\usepackage{graphicx}
\usepackage{hyperref}
\usepackage{wrapfig}
\usepackage{times}

\usepackage{algorithm}
\usepackage{algorithmic}
\usepackage{multirow}
\usepackage{rotating}
\usepackage{subfigure}
\usepackage{color}
\usepackage{enumerate}

\newtheorem{theorem}{Theorem}

\newtheorem{corollary}[theorem]{Corollary}

\newtheorem{definition}[theorem]{Definition}

\newtheorem{lemma}[theorem]{Lemma}

\newtheorem{problem}[theorem]{Problem}

\newtheorem{remark}[theorem]{Remark}

\newcommand{\im}{\mathbf{j}}

\begin{document}

\title{Topology Identification of Directed Dynamical Networks via Power Spectral Analysis}

\author{Shahin Shahrampour and Victor M. Preciado,~\IEEEmembership{Member,~IEEE,} %
\thanks{S. Shahrampour and V.M. Preciado are with the Department of Electrical
and Systems Engineering at the University of Pennsylvania, Philadelphia,
PA 19104 USA. (e-mail: shahin@seas.upenn.edu; preciado@seas.upenn.edu). %
} }

\maketitle

\begin{abstract}
We address the problem of identifying the topology of an unknown weighted, directed network of LTI systems stimulated by wide-sense stationary noises of {\it unknown} power spectral densities. We propose several reconstruction algorithms based on the cross-power spectral densities of the network's response to the input noises. Our first algorithm reconstructs the Boolean structure (i.e., existence and directions of links) of a directed network from a series of dynamical responses. Moreover, we propose a second algorithm to recover the exact structure of the network (including edge weights), as well as the power spectral density of the input noises, when an eigenvalue-eigenvector pair of the connectivity matrix is known (for example, Laplacian connectivity matrices). Finally, for the particular cases of nonreciprocal networks (i.e., networks with no directed edges pointing in opposite directions) and undirected networks, we propose specialized algorithms that result in a lower computational cost.

\end{abstract}

\thispagestyle{empty} \pagestyle{empty}



\section{Introduction}

The reconstruction of networks of dynamical
systems is an important task in many realms of science and
engineering, including  biology\cite{bonneau2006inferelator,geier2007reconstructing,bansal2007infer,julius2009genetic},
physics\cite{boccaletti2007detecting,timme2007revealing,napoletani2008reconstructing}
and finance\cite{mantegna2000introduction}. In the literature, we find a wide collection of
approaches aiming to solve the network reconstruction problem. In the physics literature, we find in \cite{timme2007revealing} a method to identify a network of dynamical systems which assumes that the input of each node can be individually manipulated. In \cite{marinazzo2008kernel}, an approach based on Granger's causality \cite{granger1969investigating} and the theory of reproducing kernel Hilbert spaces is proposed. In the statistics community, the reconstruction problem is usually approached using graphical
models by associating a random variable to each node and assuming that the (vector-valued) observations
are independent and identically distributed. In this setting, Bach and Jordan \cite{bach2004learning} used the Bayesian information criterion (BIC) to estimate sparse graphs from stationary time series. The optimization community has recently proposed a collection of papers approaching the reconstruction problem \cite{julius2009genetic, napoletani2008reconstructing, candes2008enhancing}. In these papers, several optimization
problem are proposed to find the sparsest network given a priori structural information. Although the assumption
of sparsity is well justified in some applications (e.g. biological
networks), this assumptions might lead to unsuccessful topology inference
in other cases, as illustrated in \cite{gonccalves2008necessary}
and \cite{yuan2011robust}. For tree networks, several techniques for reconstruction were proposed
in \cite{mantegna2000introduction}, \cite{materassi2009unveiling}
and \cite{materassi2010topological}. More recently, Materassi and
Salapaka proposed in \cite{materassi2012problem} a methodology for
reconstruction of directed networks using locality properties of the Wiener filters. In \cite{nabi2012sieve,6203379}, Nabi-Abdolyousefi and Mesbahi proposed techniques to extract structural information of an undirected network running consensus dynamics.

In this paper, we propose several algorithms to reconstruct the structure
of a directed network interconnecting a collection of linear dynamical systems. We first propose an algorithm to find the Boolean structure of the unknown topology. This algorithm is based on the analysis of power spectral properties of the network response when the inputs are wide-sense stationary (WSS)  processes of {\it unknown} power spectral density (PSD). Apart from recovering the Boolean structure of the network, we propose another algorithm to recover the exact structure of the network (including edge weights) when an eigenvalue-eigenvector pair of the connectivity matrix is known. This algorithm can be applied, for example, in the case of the connectivity matrix being a Laplacian matrix or the adjacency of a regular graph. Apart from general directed networks, we also propose reconstruction methodologies for directed nonreciprocal
networks (networks with no directed edges pointing in opposite directions) and undirected networks. In the latter cases, we propose specialized algorithms able to recover the network structure with less computational cost.

The rest of the paper is organized as follows. In section II, we introduce some preliminary definitions needed in our exposition and describe the network
reconstruction problem under consideration. Section III provides several theoretical results that are the foundation for our reconstruction techniques. In Section IV, we introduce several algorithms to reconstruct the Boolean structure of a directed network (Section IV.A), the exact structure of a directed network given an eigenvalue-eigenvector pair (Section IV.B), and the structure of undirected and nonreciprocal networks (Sections IV.C and IV.D, respectively). We finish with some conclusions in Section V.

\subsection*{Nomenclature}
\begin{IEEEdescription}[\IEEEusemathlabelsep\IEEEsetlabelwidth{$V_1,V_2,V_3$}]
\item[$I_{d}$] $d\times d$ identity matrix.
\item[$\mathbf{1}_{d}$] $d$-dimensional vector of all ones.
\item[$\mathbb{E}(\cdot)$] Expectation operator.
\item[$R_{xy}(\tau )$] Cross-correlation function, $\mathbb{E}(x(t)y(t-\tau ))$.
\item[$R_{x}(\tau )$] Auto-correlation function, $\mathbb{E}(x(t)x(t-\tau ))$.
\item[$\mathcal{F}\left\{ \cdot \right\} $] Fourier transform.
\item[$S_{y_{i}y_{j}}(\omega )$] Cross-power spectral density (CPSD), $\mathcal{F}\big\{R_{y_{i}y_{j}}(\tau )\big\}$.
\item[$S_{y_{i}}(\omega )$] Power spectral density (PSD), $\mathcal{F}\big\{R_{y_{i}y_{i}}(\tau )\big\}$.
\end{IEEEdescription}

\section{Preliminaries \& Problem Description}

In this section, we state the problem under consideration. First,
we introduce some notions of graph theory which are needed in our
derivations (see \cite{west2001introduction}, for an extensive exposition).

\subsection{Graph Theory}

Let $\mathcal{G}\triangleq\left(\mathcal{V},\mathcal{E}\right)$ be
an unweighted, undirected graph, where $\mathcal{V}\triangleq\left\{ v_{1},\dots,v_{N}\right\} $
denotes a set of $N$ nodes and $\mathcal{E}\subseteq\mathcal{V}\times\mathcal{V}$
denotes a set of $m$ undirected edges. If $\left\{ v_{i},v_{j}\right\} \in\mathcal{E}$,
we call nodes $v_{i}$ and $v_{j}$ \emph{adjacent} (or first-neighbors),
which we denote by $v_{i}\sim v_{j}$.\ A \emph{weighted}, undirected
graph is defined as the triad $\mathcal{W}\triangleq\left(\mathcal{V},\mathcal{E},\mathcal{F}\right)$,
where $\mathcal{V}$ and $\mathcal{E}$ are the sets of nodes and
edges in $\mathcal{W}$, and the function $\mathcal{F}:\mathcal{E\rightarrow}\mathbb{R}$
associates real weights to the edges. Similarly, a weighted, \emph{directed}
graph is defined as the triad $\mathcal{D}\triangleq\left(\mathcal{V},\mathcal{E}_{d},\mathcal{F}_{d}\right)$,
where $\mathcal{V}$ is the set of nodes and $\mathcal{E}_{d}$ is
the set of directed edges in $\mathcal{D}$, where a directed edge
from node $v_{i}$ to node $v_{j}$ is defined as an ordered pair
$\left(v_{i},v_{j}\right)$. Furthermore, $\mathcal{F}_{d}$ is a
weight function $\mathcal{F}_{d}:\mathcal{E}_{d}\rightarrow\mathbb{R}$.

In an unweighted, undirected graph $\mathcal{G}$, the \emph{degree}
of a vertex $v_{i}$, denoted by $\deg\left(v_{i}\right)$, is the
number of nodes adjacent to it, i.e., $\deg\left(v_{i}\right)=\left\vert \left\{ v_{j}\in\mathcal{V}:\left\{ v_{i},v_{j}\right\} \in\mathcal{E}\right\} \right\vert $.
This definition can be generalized to both weighted and directed graphs.
For weighted graphs, the weighted degree of node $v_{i}$ is equal
to $\deg\left(v_{i}\right)=\sum_{j:\left\{ v_{i},v_{j}\right\} \in\mathcal{E}}\mathcal{F}\left(\left\{ v_{i},v_{j}\right\} \right)$,
i.e., the sum of the weights associated to edges connected to $v_{i}$.
For weighted, directed networks, we define the weighted \emph{in-degree}
of node $v_{i}$ as $\deg_{in}\left(v_{i}\right)=\sum_{j:\left(v_{j},v_{i}\right)\in\mathcal{E}_{d}}\mathcal{F}_{d}\left(\left(v_{j},v_{i}\right)\right)$.

The \emph{adjacency matrix} of an unweighted, undirected graph $\mathcal{G}$,
denoted by $A_{\mathcal{G}}=[a_{ij}]$, is a $N\times N$ Boolean
symmetric matrix defined entry-wise as $a_{ij}=1$ if nodes $v_{i}$
and $v_{j}$ are adjacent, and $a_{ij}=0$ otherwise. We define the
\emph{Laplacian matrix }$L_{\mathcal{G}}$ of a graph $\mathcal{G}$
as $L_{\mathcal{G}}=D_{\mathcal{G}}-A_{\mathcal{G}}$ where $D_{\mathcal{G}}$
is the diagonal matrix of degrees, $D_{\mathcal{G}}=diag\left(\left(\deg\left(v_{i}\right)\right)_{i=1}^{N}\right)$.
For simple graphs, $L_{\mathcal{G}}$ is a symmetric positive semidefinite
matrix, which we denote by $L_{\mathcal{G}}\succeq0$ \cite{biggs1993algebraic}.
Thus, $L_{\mathcal{G}}$ has a full set of $N$ real and orthogonal
eigenvectors with real nonnegative eigenvalues $0=\lambda_{1}\leq\lambda_{2}\leq...\leq\lambda_{N}$.

Similarly, the weighted adjacency matrix of a weighted graph $\mathcal{W}$
is defined as $A_{\mathcal{W}}=\left[w_{ij}\right]$, where $w_{ij}=\mathcal{F}\left(\left\{ v_{i},v_{j}\right\} \right)$
for $\left\{ v_{i},v_{j}\right\} \in\mathcal{E}$, and $w_{ij}=0$
if $\left\{ v_{i},v_{j}\right\} \not\in\mathcal{E}$. We define the
\emph{degree matrix} of a weighted graph $\mathcal{W}$ as the diagonal
matrix $D_{\mathcal{W}}=diag\left(\left(\deg\left(v_{i}\right)\right)_{i=1}^{N}\right)$.
The Laplacian matrix of a weighted, undirected graph $\mathcal{W}$,
is defined as $L_{\mathcal{W}}=D_{\mathcal{W}}-A_{\mathcal{W}}$.
Furthermore, the adjacency matrix of a weighted, directed graph $\mathcal{D}$
is defined as $A_{\mathcal{D}}=\left[d_{ij}\right]$, where $d_{ij}=\mathcal{F}_{d}\left(\left(v_{j},v_{i}\right)\right)$
for $\left(v_{j},v_{i}\right)\in\mathcal{E}_{d}$, and $d_{ij}=0$
if $\left(v_{j},v_{i}\right)\not\in\mathcal{E}_{d}$. We define the
\emph{in-degree matrix} of a directed graph $\mathcal{D}$ as the
diagonal matrix $D_{\mathcal{D}}=diag\left(\left(\deg_{in}\left(v_{i}\right)\right)_{i=1}^{N}\right)$.
The Laplacian matrix of $\mathcal{D}$ is then defined as $L_{\mathcal{D}}=D_{\mathcal{D}}-A_{\mathcal{D}}$.
The Laplacian matrix, for all the unweighted, weighted, and directed
cases, satisfies $L_{\mathcal{G}}\mathbf{1}=L_{\mathcal{W}}\mathbf{1}=L_{\mathcal{D}}\mathbf{1}=\mathbf{0}$,
i.e., the vector $\mathbf{1}/\sqrt{N}$ is an eigenvector of the Laplacian
matrix with eigenvalue $0$.

\subsection{Dynamical Network Model \& Problem Statement}

Consider a dynamical network consisting of $N$ linearly coupled identical
nodes, with each node being an $n$-dimensional LTI SISO dynamical
system. The dynamical network under study can be characterized by 
\begin{align}
\dot{x}_{i}(t) & =Ax_{i}(t)+b\left(\sum_{j=1}^{N}g_{ij}y_{j}(t)+w_{i}\left(t\right)\right),\label{dynamics}\\
y_{i}(t) & =c^{T}x_{i}(t),\nonumber 
\end{align}
where $x_{i}(t)\in\mathbb{R}^{n}$ denotes the state vector describing
the dynamics of node $v_{i}\in\mathcal{V}$. $A\in\mathbb{R}^{n\times n}$
and $b,c\in\mathbb{R}^{n}$ are the given state, input and output
matrices corresponding to the state-space representation of each node
in isolation. ${w}_{i}\left(t\right)$ and $y_{i}(t)\in\mathbb{R}$
are stochastic processes representing the input noise and the system
output, respectively. $g_{ij}\geq0$ is the coupling strength of a
\textit{directed} edge from $v_{i}$ to $v_{j}$, which we shall assume
to be \textit{unknown}.

Defining the network state vector $\mathbf{x}(t)\triangleq(x_{1}^{T}(t),\ldots,x_{N}^{T}(t))^{T}\in\mathbb{R}^{Nn}$,
the noise vector $\mathbf{w}(t)\triangleq(w_{1}(t),\ldots,w_{N}(t))^{T}\in\mathbb{R}^{N}$,
and the network output vector $\mathbf{y}(t)\triangleq(y_{1}(t),\ldots,y_{N}(t))^{T}\in\mathbb{R}^{N}$,
respectively, we can rewrite the network dynamics in \eqref{dynamics},
as 
\begin{align}
\dot{\mathbf{x}}(t) & =\left(I_{N}\otimes A+\mathbf{G}\otimes bc^{T}\right)\mathbf{x}(t)+\left(I_{N}\otimes b\right)\mathbf{w}(t),\label{dynamics2}\\
\mathbf{y}(t) & =\left(I_{N}\otimes c^{T}\right)\mathbf{x}(t),\nonumber 
\end{align}
where $\mathbf{G}=[g_{ij}]$ is the \textit{connectivity} matrix of
a (possibly weighted and/or directed) network $\mathcal{D}$. For
the networked dynamical system to be stable, we assume the network
state matrix $I_{N}\otimes A+\mathbf{G}\otimes bc^{T}$ to be Hurwitz.

Hereafter, we will analyze the following scenario. Consider a collection
of $N$ dynamical nodes with a known LTI SISO dynamics defined by
the state-space matrices $(A,b,c^{T},0)$. The link structure of the
network dynamic model, described by the connectivity matrix $\mathbf{G}$,
is completely unknown. We assume the input noises, ${w_{i}\left(t\right)}$,
are i.i.d. wide-sense stationary processes of \textit{unknown} but
identical power spectral densities, i.e., $S_{w_{i}}(\omega)=S_{w}(\omega)$
for all $i=1,\ldots,N$. We are interested in identifying all the
links in the network by exploiting only the information provided by
the realizations of the output stochastic processes $y_{1}(t),\ldots,y_{N}(t)$.
Formally, we can formulate this problem as follows:
\begin{problem}
\label{Main Problem}Consider the dynamical network model in \eqref{dynamics2},
whose connectivity matrix $\mathbf{G}$ is unknown. Assume that the
only available information is a spectral characterization of the
output signals $y_{1}(t),\ldots,y_{N}(t)$ in terms of power and cross-power
spectral densities, $S_{y_{i}}(\omega)$ and $S_{y_{i}y_{j}}(\omega)$,
which can be empirically estimated from the output signals\footnote{One can use, for example, BartlettÕs averaging method \cite{brillinger1981time} to produce periodogram
estimates of power and cross-power spectral densities, $S_{y_{i}}(\omega)$ and $S_{y_{i}y_{j}}(\omega)$.}. Then, find
the Boolean structure of the directed network, i.e., the location and
directions of all the edges. 
\end{problem}
It is worth remarking that we assume the input noise to be an exogenous
signal of \textit{unknown} power spectral density, $S_{w}(\omega)$.
We will provide in Section \ref{BooleanReconstruction} a methodology
to recover the Boolean structure of the network, even though the input
noise is not known. We will show in Section \ref{ExactReconstruction}
that for certain connectivity matrices, such as Laplacian matrices,
we can recover the weights of the directed edges in the network, as
well as the power spectral density of the input noise. Moreover, in Sections \ref{Undirected Reconstruction} and \ref{NonReciprocal Reconstruction} we provide two computationally efficient algorithms to recover the structure of undirected and nonreciprocal networks, respectively.

\section{Theoretical Results\label{main}}

We start by stating some assumptions we need in our subsequent developments.
The following definition will be useful for determining sufficient
conditions for detection of links in a network.

\begin{definition}{[}\textit{Excitation Frequency Interval, \cite{materassi2012problem}}{]}
The excitation frequency interval of a vector $\mathbf{w}\left(t\right)$
of wide-sense stationary processes is defined as an interval $(-\Omega,\Omega)$,
with $\Omega>0$, such that the power spectral densities of the input components
$w_{i}\left(t\right)$ satisfy $S_{w_{i}}(\omega)>0$ for all $\omega\in(-\Omega,\Omega)$,
and all $i\in\{1,2,...,N\}$.\\
 \end{definition}
 
 Throughout the paper we impose the following
conditions on the input vector:
 
\begin{description}
\item [{A1.}] The collection of signals $\left\{ w_{i}(t),i=1,...,N\right\} $
are uncorrelated zero-mean WSS processes with identical autocorrelation
function, i.e., for any $t,\tau\in\mathbb{R}$, $R_{w_{i}}(\tau)=\mathbb{E}(w_{i}(t)w_{i}(t+\tau))\triangleq R_{w}(\tau)$. 
\item [{A2.}] The input noise $\mathbf{w}\left(t\right)$ presents a nonempty
excitation frequency interval $\left(-\Omega,\Omega\right)$. 
\end{description}
In our derivations, we will invoke the following variation of the
matrix inversion lemma \cite{TS86}:

\begin{lemma}[Sherman-Morrison-Woodbury]\label{inversion lemma} Assume that the matrices $D$ and
$I+WD^{-1}UE$ are nonsingular. Then, the following identity holds
\begin{align*}
\left(D+UEW\right)^{-1}=D^{-1}-D^{-1}UE\left(I+WD^{-1}UE\right)^{-1}WD^{-1},
\end{align*}
where $E,W,D,$ and $U$ are matrices of compatible dimensions and
$I$ is the identity matrix. \end{lemma}

Based on Woodbury's formula, we derive an expression that provides
an explicit relationship between the (cross-)power spectral densities
of two stochastic outputs, $y_{i}\left(t\right)$ and $y_{j}\left(t\right)$,
when we inject a noise $w_{k}\left(t\right)$ into node $k$ with
power spectral density $S_{w}\left(\omega\right)$.

\begin{lemma} \label{Power Spectral Densities}Consider the continuous-time
networked dynamical system \eqref{dynamics2}. Then, under assumptions
(A1)-(A2), the following identity holds 
\begin{equation}
\mathbf{S}\left(\omega\right)=S_{w}(\omega)\left(\frac{I_{N}}{\left|h\left(\im\omega\right)\right|^{2}}+\mathbf{G}^{T}\mathbf{G}-\frac{\mathbf{G}}{h^{\ast}\left(\im\omega\right)}-\frac{\mathbf{G}^{T}}{h\left(\im\omega\right)}\right)^{-1},
\label{spectrum}
\end{equation}
where $\mathbf{S}\left(\omega\right)\triangleq\left[S_{y_{i}y_{j}}(\omega)\right]$
is the matrix of output CPSD's, and $h\left(\im\omega\right)\triangleq c^{T}\left(\im\omega I_{n}-A\right)^{-1}b$
is the nodal transfer function.
\end{lemma}

\begin{IEEEproof} The $N\times N$ transfer matrix, $H\left(\im w\right)\triangleq\left[H_{ji}\left(\im\omega\right)\right]$,
of the state-space model in \eqref{dynamics2} is given by 
\begin{align}
H\left(\im\omega\right) & =(I_{N}\otimes c^{T})\bigg(\im\omega I_{Nn}-I_{N}\otimes A-\mathbf{G}\otimes bc^{T}\bigg)^{-1}(I_{N}\otimes b)\nonumber \\
 & =(I_{N}\otimes c^{T})\bigg(I_{N}\otimes(\im\omega I_{n}-A)-\mathbf{G}\otimes bc^{T}\bigg)^{-1}(I_{N}\otimes b).\label{trfunction}
\end{align}
Assume we inject a noise signal into the $k$-th node, i.e., $\mathbf{w}\left(t\right)=w_{k}\left(t\right)\mathbf{e}_{k}$.
Hence, the power spectral density measured on the output of node $i$
is equal to $S_{y_{i}}(\omega)=H_{ki}(\omega)H_{ki}^{\ast}(\omega)S_{w_{k}}(\omega)$.
On the other hand, the transfer functions from input $w_{k}\left(t\right)$
to the outputs $y_{i}\left(t\right)$ and $y_{j}\left(t\right)$ are,
respectively, $Y_{i}\left(\im\omega\right)/W_{k}\left(\im\omega\right)=H_{ki}(\im\omega)$
and $Y_{j}\left(\im\omega\right)/W_{k}\left(\im\omega\right)=H_{kj}(\im\omega)$,
where $Y_{i}\left(\im\omega\right)$ and $W_{k}\left(\im\omega\right)$
are the Fourier transforms of $y_{i}\left(t\right)$ and $w_{k}\left(t\right)$,
respectively. Hence, $Y_{j}\left(\im\omega\right)/Y_{i}\left(\im\omega\right)=H_{ki}^{-1}(\im\omega)H_{kj}(\im\omega)$
which implies $S_{y_{i}y_{j}}(\omega)=\bigg(H_{kj}(\im\omega)H_{ki}^{-1}(\im\omega)\bigg)^{\ast}S_{y_{i}}(\omega)$.
Since $S_{w_{k}}(\omega)=S_{w}(\omega)$ for all $k$, we have that
$S_{y_{i}y_{j}}(\omega)=H_{ki}(\im\omega)H_{kj}^{\ast}(\im\omega)S_{w}(\omega)$.

Assume that we inject noise signals satisfying assumptions (A1)-(A2)
into all the nodes in the network, i.e., $\mathbf{w}\left(t\right)=\sum_{k=1}^{N}w_{k}\left(t\right)\mathbf{e}_{k}$.
Hence, we can apply superposition to obtain 
\begin{align}
\frac{S_{y_{i}y_{j}}(\omega)}{S_{w}(\omega)} & =\sum_{k=1}^{N}H_{kj}^{\ast}(\im\omega)H_{ki}(\im\omega)\nonumber \\
 & =\sum_{k=1}^{N}\mathbf{e}_{k}^{T}H^{\ast}\left(\im\omega\right)\mathbf{e}_{j}\mathbf{e}_{i}^{T}H\left(\im\omega\right)\mathbf{e}_{k}\nonumber \\
 & =\sum_{k=1}^{N}\text{Tr}\bigg(H^{\ast}\left(\im\omega\right)\mathbf{e}_{j}\mathbf{e}_{i}^{T}H\left(\im\omega\right)\mathbf{e}_{k}\mathbf{e}_{k}^{T}\bigg)\nonumber \\
 & =\text{Tr}\bigg(H^{\ast}\left(\im\omega\right)\mathbf{e}_{j}\mathbf{e}_{i}^{T}H\left(\im\omega\right)\sum_{k=1}^{N}\mathbf{e}_{k}\mathbf{e}_{k}^{T}\bigg)\nonumber \\
 & =\mathbf{e}_{i}^{T}H\left(\im\omega\right)H^{\ast}\left(\im\omega\right)\mathbf{e}_{j},\label{product}
\end{align}
for any $\omega\in\left(-\Omega,\Omega\right)$, where we used the
identity $\sum_{k=1}^{N}\mathbf{e}_{k}\mathbf{e}_{k}^{T}=I_{N}$ in
our derivation. 

Let us define the matrices $W\triangleq I_{N}\otimes c^{T}$, $U\triangleq I_{N}\otimes b$,
$E\triangleq-\mathbf{G}$, and $D\triangleq I_{N}\otimes\left(\im\omega I_{n}-A\right)$.
Then, we can rewrite the transfer matrix $H\left(\im\omega\right)$
in \eqref{trfunction} as 
\begin{align}
H\left(\im\omega\right)=W(D+UEW)^{-1}U.\label{product2}
\end{align}
Also, we have that $h\left(\im\omega\right)I_{N}=WD^{-1}U$. Then,
applying Lemma \ref{inversion lemma} to \eqref{product2}, we can
rewrite the transfer matrix, as follows 
\begin{align*}
H\left(\im\omega\right) & =h\left(\im\omega\right)\bigg(I_{N}+\mathbf{G}\big(I_{N}-h\left(\im\omega\right)\mathbf{G}\big)^{-1}h\left(\im\omega\right)I_{N}\bigg)\\
 & =h\left(\im\omega\right)\bigg(I_{N}+\mathbf{G}\big(\frac{I_{N}}{h\left(\im\omega\right)}-\mathbf{G}\big)^{-1}\bigg)\\
 & =h\left(\im\omega\right)\bigg(I_{N}+\big(\mathbf{G}-\frac{I_{N}}{h\left(\im\omega\right)}+\frac{I_{N}}{h\left(\im\omega\right)}\big)\big(\frac{I_{N}}{h\left(\im\omega\right)}-\mathbf{G}\big)^{-1}\bigg)\\
 & =h\left(\im\omega\right)\bigg(I_{N}-I_{N}+\frac{1}{h\left(\im\omega\right)}\big(\frac{I_{N}}{h\left(\im\omega\right)}-\mathbf{G}\big)^{-1}\bigg)\\
 & =\big(\frac{I_{N}}{h\left(\im\omega\right)}-\mathbf{G}\big)^{-1}
\end{align*}
Substituting the above into \eqref{product}, we reach the statement of our lemma after a simple expansion of the resulting expression.

\end{IEEEproof}

In the following section,
we will use this lemma to reconstruct an unknown network structure
$\mathbf{G}$ from the empirical CPSD's of the outputs. We will
also show that, assuming that we know one eigenvalue-eigenvector pair
of $\mathbf{G}$, we can recover the weighted and directed graph $\mathcal{D}$
(not only its Boolean structure), as well as the PSD of the noise,
$S_{w}\left(\omega\right)$. Relevant examples of this situation are:
(\emph{i}) networks of diffusively coupled systems with a Laplacian
connectivity matrix \cite{shahrampour2013reconstruction}, i.e., $\mathbf{G}=-L_{\mathcal{D}}$,
since Laplacian matrices always satisfy $L_{\mathcal{D}}\mathbf{1}_{N}=0$;
or (\emph{ii}) $k$-regular networks \cite{west2001introduction},
i.e., $\mathbf{G}=A_{k}$, since the adjacency matrix $A_{k}$ satisfy
$A_{k}\boldsymbol{1}_{N}=k$.

As stated in Problem \ref{Main Problem}, the PSD of the input noise
$\mathbf{w}\left(t\right)$ is not available to us to perform the
network reconstruction. The following lemma will allow us reconstruct
this PSD when an eigenvalue-eigenvector pair of $\mathbf{G}$ is known
\emph{a priori}.

\begin{lemma}\label{input spectrum} Consider the continuous-time
networked dynamical system \eqref{dynamics2}. Then, under assumptions
(A1)-(A2), the input PSD can be computed as 
\begin{align}
S_{w}(\omega) & =\frac{\lambda^{2}|h\left(\im\omega\right)|^{2}-2\lambda\text{Re}\{h\left(\im\omega\right)\}+1}{(\boldsymbol{u}^{T}\mathbf{S}^{-1}\left(\omega\right)\boldsymbol{u})|h\left(\im\omega\right)|^{2}},\label{input power spectrum}
\end{align}
where $\left(\lambda,\boldsymbol{u}\right)$ is an eigenvalue-eigenvector
pair of $\mathbf{G}$, $h\left(\im\omega\right)$ is the nodal transfer
function, and $\mathbf{S}\left(\omega\right)\triangleq\left[S_{y_{i}y_{j}}(\omega)\right]$
is the matrix of CPSD's.

\end{lemma}

\begin{IEEEproof} From \eqref{spectrum}, we have 
\begin{align*}
\mathbf{S}^{-1}\left(\omega\right)S_{w}(\omega)=\frac{I_{N}}{\left|h\left(\im\omega\right)\right|^{2}}+G^{T}G-\frac{G}{h^{\ast}\left(\im\omega\right)}-\frac{G^{T}}{h\left(\im\omega\right)}.
\end{align*}
Pre- and post-multiplying by $\boldsymbol{u}^{T}$ and $\boldsymbol{u}$,
respectively, we obtain 
\begin{align*}
\left(\boldsymbol{u}^{T}\mathbf{S}^{-1}\left(\omega\right)\boldsymbol{u}\right)S_{w}(\omega)=\frac{1}{\left|h\left(\im\omega\right)\right|^{2}}+\lambda^{2}-\frac{\lambda}{h\left(\im\omega\right)}-\frac{\lambda}{h^{\ast}\left(\im\omega\right)}.
\end{align*}
Dividing by $\boldsymbol{u}^{T}\mathbf{S}^{-1}\left(\omega\right)\boldsymbol{u}$,
we reach \eqref{input power spectrum}.

\end{IEEEproof}

Lemma \ref{input spectrum} shows that, given the eigenvalue-eigenvector pair $\left(\lambda,\boldsymbol{u}\right)$,
the PSD of the input noise can be reconstructed from the nodal
transfer function and the matrix of CPSD's, $\mathbf{S}\left(\omega\right)$,
which can be numerically approximated from the empirical cross-correlations between
output signals.

\section{Reconstruction Methodologies\label{Methodologies}}

Based on the above results, we introduce several methodologies to
reconstruct the structure of an unknown network following the dynamics
in \eqref{dynamics2} when \emph{the
PSD of the input noise is unknown}. First, in Subsection \ref{BooleanReconstruction},
we present a technique to reconstruct the Boolean structure of an
unknown (possibly weighted) directed network. Moreover, if an eigenvalue-eigenvector
pair of $\mathbf{G}$ is known (for example, $\mathbf{G}$ is a Laplacian
matrix), we show how to recover the weights of the directed edges,
as well as the PSD of the input noise in Subsection \ref{ExactReconstruction}.
Finally, in Subsections \ref{Undirected Reconstruction} and \ref{NonReciprocal Reconstruction}, we provide
reconstruction techniques to recover two special cases, namely, undirected
networks and nonreciprocal directed networks, respectively.

Consider Problem \ref{Main Problem}, when $\mathbf{G}$ is an unknown
connectivity matrix representing a weighted, directed network $\mathcal{D}$.
We propose a reconstruction technique to recover the Boolean structure
of $\mathcal{D}$ when the PSD of the input noise is \emph{unknown}.
Note that, in general, the result in Lemma \ref{Power Spectral Densities}
is not enough to extract the underlying structure of the network,
even if the input noise PSD were known. In what follows, we propose
a methodology to reconstruct a directed network of dynamical nodes
by {\it grounding} the dynamics in a series of nodes, similar to
the approach proposed in \cite{6203379} to reconstruct undirected
networks following a consensus dynamics.

\begin{definition}[Grounded Dynamics]\label{groundedcon} The dynamics
of \eqref{dynamics2} grounded at node $v_{j}$ takes the form 
\begin{align}
\mathbf{\dot{\widetilde{x}}}\left(t\right) & =\left(I_{N-1}\otimes A+\mathbf{\widetilde{G}}_{j}\otimes bc^{T}\right)\widetilde{\mathbf{x}}(t)+\left(I_{N-1}\otimes b\right)\widetilde{\mathbf{w}}(t),\label{groundedlaplacian}\\
\mathbf{\widetilde{y}}(t) & =\left(I_{N-1}\otimes c^{T}\right)\widetilde{\mathbf{x}}(t),\nonumber 
\end{align}
where $\widetilde{\mathbf{w}}(t)$ is obtained by eliminating the
$j$-th entry from the noise input $\mathbf{w}\left(t\right)$, and
$\mathbf{\widetilde{G}}_{j}\in\mathbb{R}^{(N-1)\times(N-1)}$ is obtained
by eliminating the $j$-th row and column from $\mathbf{G}$.

\end{definition}

The dynamics in \eqref{groundedlaplacian} describes the evolution
of \eqref{dynamics2} when we ground the state of node $v_{j}$
to be $x_{j}(t)\equiv0$. Applying Lemma \ref{Power Spectral Densities}
to the grounded dynamics \eqref{groundedlaplacian}, one obtains
the following expression for the CPSD's:

\begin{equation}
\widetilde{\mathbf{S}}_{j}(\omega)=S_{w}(\omega)\left(\frac{I_{N}}{\left|h\left(\im\omega\right)\right|^{2}}+\widetilde{\mathbf{G}}_{j}^{T}\widetilde{\mathbf{G}}_{j}-\frac{\widetilde{\mathbf{G}}_{j}}{h^{\ast}\left(\im\omega\right)}-\frac{\widetilde{\mathbf{G}}_{j}^{T}}{h\left(\im\omega\right)}\right)^{-1}.\label{spectrum2}
\end{equation}

We will use the next Theorem to propose several reconstruction techniques
in Subsections \ref{BooleanReconstruction} and \ref{ExactReconstruction}.

\begin{theorem}\label{Directed Network} 

Consider the networked dynamical system \eqref{dynamics2} with connectivity
matrix $\mathbf{G}=\left[g_{ij}\right]$. Let us denote by $S_{w}\left(\omega\right)$
the PSD of the input noise, by $\mathbf{S}\left(\omega\right)=\left[S_{y_{i}y_{j}}(\omega)\right]$
the $N\times N$ matrix of CPSD's for the (ungrounded) dynamics \eqref{dynamics2},
and by $\mathbf{\widetilde{S}}_{j}\left(\omega\right)=[\widetilde{S}_{y_{i}y_{k}}(\omega)]_{i,k\neq j}$
the $N-1\times N-1$ matrix of CPSD's for the dynamics in \eqref{groundedlaplacian}
grounded at node $v_{j}$. Then, under assumptions (A1)-(A2), we have
that, for any $\omega_{0}\in\left(-\Omega,\Omega\right)$,

\begin{equation}
g_{ji}=\left\{ \begin{array}{ll}
\left[S_{w}\left(\omega_{0}\right)\left([\mathbf{S}^{-1}\left(\omega_{0}\right)]_{ii}-[\mathbf{\mathbf{\widetilde{S}}}_{j}^{-1}\left(\omega_{0}\right)]_{ii}\right)\right]^{1/2}, & \text{for }i<j,\\
\left[S_{w}\left(\omega_{0}\right)\left([\mathbf{S}^{-1}\left(\omega_{0}\right)]_{ii}-[\mathbf{\widetilde{S}}_{j}^{-1}\left(\omega_{0}\right)]_{i-1,i-1}\right)\right]^{1/2}, & \text{for }i>j.
\end{array}\right.\label{G entry}
\end{equation}
\end{theorem} 
\begin{IEEEproof}
Without loss of generality, we consider the case $j=N$ (for any other
$j\neq N$, we can transform the problem to the case $j=N$ via a
simple reordering of rows and columns). Subtracting the diagonal elements
of $\mathbf{S}^{-1}\left(\omega\right)$ in \eqref{spectrum2} from
those of $\mathbf{\widetilde{S}}_{j}^{-1}\left(\omega\right)$ in
\eqref{spectrum}, we obtain 
\begin{align*}
[\mathbf{S}^{-1}\left(\omega\right)]_{ii}-[\mathbf{\widetilde{S}}_{j}^{-1}\left(\omega\right)]_{ii} & =\frac{[\mathbf{G}^{T}\mathbf{G}]_{ii}-[\mathbf{\widetilde{G}}_{N}^{T}\widetilde{\mathbf{G}}_{N}]_{ii}}{S_{w}(\omega)}.
\end{align*}
Also, since $[\mathbf{G}^{T}\mathbf{G}]_{ii}=\sum_{k}g_{ki}^{2}$
and $[\mathbf{\widetilde{G}}_{N}^{T}\widetilde{\mathbf{G}}_{N}]_{ii}=\sum_{k\neq N}g_{ki}^{2}$,
we have that 
\begin{align*}
[\mathbf{G}^{T}\mathbf{G}]_{ii}-[\mathbf{\widetilde{G}}_{N}^{T}\widetilde{\mathbf{G}}_{N}]_{ii}=g_{Ni}^{2},
\end{align*}
for any $i<N$. The same analysis holds for
$j\neq N$. Hence, we can recover the entries $g_{ji}$, for $i<j$, as
stated in our Theorem. Notice also that, for $j\neq N$ and $i>j$, we must use the
entry $[\mathbf{\widetilde{S}}_{j}^{-1}\left(\omega\right)]_{i-1,i-1}$
in \eqref{G entry}, to take into account that $\mathbf{\widetilde{S}}_{j}\left(\omega\right)$
is an $(N-1)\times (N-1)$ matrix associated to the dynamics grounded at node $v_j$.
\end{IEEEproof}

\subsection{Boolean Reconstruction of Directed Networks\label{BooleanReconstruction}}

Theorem \ref{Directed Network} allows us to reconstruct the Boolean
structure of an unknown directed network if we have access to the matrices
of CPSD's, $\mathbf{S}\left(\omega_{0}\right)$ and $\mathbf{\mathbf{\widetilde{S}}}_{j}\left(\omega_{0}\right)$,
for any $\omega_{0}$ in the excitation frequency interval $\left(-\Omega,\Omega\right)$.
In particular, one can verify the existence of a directed edge $\left(i,j\right)$
by checking the condition $g_{ji}>0$, where $g_{ji}$ is computed
from \eqref{G entry}. In practice, the CPSD's $\mathbf{S}\left(\omega_{0}\right)$
and $\mathbf{\mathbf{\widetilde{S}}}_{j}\left(\omega_{0}\right)$ are
empirically computed from the stochastic outputs of the network, $\mathbf{y}\left(t\right)$
and $\widetilde{\mathbf{y}}\left(t\right)$; therefore, they are subject
to numerical errors. Hence, in the implementation, one should
relax the condition $g_{ji}>0$ to $g_{ji}>\tau$, where $\tau$ is
a small threshold used to account for numerical precision.

Based on Theorem \ref{Directed Network}, we propose Algorithm \ref{BooleanAlg} to find the Boolean representation of $\mathbf{G}$, denoted
by $\mathbf{B}\left(\mathbf{G}\right)$, when a directed dynamical
network is excited by an input noise of \emph{unknown PSD}.

\begin{algorithm}[t]
\caption{Boolean reconstruction of directed networks} \label{BooleanAlg}
\begin{algorithmic}[1]
\REQUIRE $h(\im \omega)$, $\mathbf{y}(t)$ from \eqref{dynamics2}, $\widetilde{\mathbf{y}}(t)$ from \eqref{groundedlaplacian}, and any $\omega_{0} \in (-\Omega,\Omega)$;%
\STATE Compute $\mathbf{S}(\omega_{0})$ from $\mathbf{y(t)}$;
\FOR{$j=1:N$}%
\STATE Compute $\mathbf{\widetilde{S}}_{j}(\omega_{0})$ from $\widetilde{\mathbf{y}}(t)$;
\FOR{$i=1:j-1$}
\STATE \bf{if} $[\mathbf{S}^{-1}\left(\omega_{0}\right)]_{ii}-[\mathbf{\widetilde{S}}_{j}^{-1}\left(\omega_{0}\right)]_{ii}>\tau$ then $b_{ji}=1$;%
\STATE \bf{if} $[\mathbf{S}^{-1}\left(\omega_{0}\right)]_{ii}-[\mathbf{\widetilde{S}}_{j}^{-1}\left(\omega_{0}\right)]_{ii}<\tau$ then $b_{ji}=0$;%
\ENDFOR
\FOR{$i=j+1:N$}
\STATE \bf{if} $[\mathbf{S}^{-1}\left(\omega_{0}\right)]_{ii}-[\mathbf{\widetilde{S}}_{j}^{-1}\left(\omega_{0}\right)]_{i-,1i-1}>\tau$ then $b_{ji}=1$;%
\STATE \bf{if} $[\mathbf{S}^{-1}\left(\omega_{0}\right)]_{ii}-[\mathbf{\widetilde{S}}_{j}^{-1}\left(\omega_{0}\right)]_{i-1,i-1}<\tau$ then $b_{ji}=0$;%
\ENDFOR
\ENDFOR
\end{algorithmic}
\end{algorithm}

Algorithm \ref{BooleanAlg} incurs the following computational cost:

\begin{enumerate}[\itshape i)]
\item It computes the cross-correlation functions for all the
$N^2$ pairs of outputs in \eqref{dynamics2}. For each one of the $N$ grounded dynamics in \eqref{groundedlaplacian}, the algorithm also computes $N^2$ pairs of cross-correlation functions, resulting in a total of $N^3$. To compute these cross-correlations
we use time series of length $L$. Since each each cross-correlation
takes $\mathcal{O}\left(L^{2}\right)$ operations, we have a total
of $\mathcal{O}\left(N^{3}L^{2}\right)$ operations to compute all
the required cross-correlations.

\item Algorithm \ref{BooleanAlg} computes the FFT of all the $\left(N+1\right)N^2$
cross-correlation function of length $L$ in (\emph{i}) at a particular
frequency $\omega_{0}\in(-\Omega,\Omega)$. Since computing the FFT
at a single frequency takes $\mathcal{O}(L)$ operations, we have
a total of $\mathcal{O}\left(N^{3}L\right)$ operations to compute
the CPSD's matrices $\mathbf{S}\left(\omega_{0}\right)$ and $\mathbf{\mathbf{\widetilde{S}}}_{j}\left(\omega_{0}\right)$,
for all $j=1,\ldots,N$.

\item Our algorithm also needs to compute the inverse of $\mathbf{S}\left(\omega\right)$
and $\mathbf{\mathbf{\widetilde{S}}}_{j}\left(\omega\right)$. Since
each inversion takes $\mathcal{O}\left(N^3\right)$, we have
a total of $\mathcal{O}\left(N^4\right)$ operations to compute the inverses
of all the $N+1$ matrices involved in our computations.

\end{enumerate}

Therefore, the total computational cost of our algorithm is $\mathcal{O}\left(N^4+N^{3}L^{2}\right)$.
In the next subsection, we extend Algorithm \ref{BooleanAlg} to reconstruct
the exact connectivity matrix $\mathbf{G}$.

\subsection{Exact Reconstruction of Directed Networks\label{ExactReconstruction}}

Apart from a Boolean reconstruction of $\mathbf{G}$, we can also
compute the weights of the edges in the network if we know one eigenvalue-eigenvector
pair $(\lambda,\mathbf{u})$ of $\mathcal{\mathbf{G}}$, as follows.
This is the case of $\mathcal{\mathbf{G}}$ being, for example, a
Laplacian matrix (since $\mathbf{G}\mathbf{1}_{N}=0$, in this case),
or the adjacency matrix of a $d$-regular graph (since $\mathbf{G}\mathbf{1}_{N}=d\mathbf{1}_{N}$).
In these cases, we use Lemma \ref{input power spectrum} to find the
value of $S_{w}\left(\omega_{0}\right)$ at a particular frequency
$\omega_{0}\in(-\Omega,\Omega)$. For example, in the case of $\mathbf{G}$
being a Laplacian, we have the following result:

\begin{corollary}\label{input spectrum2} Consider the networked
dynamical system in \eqref{dynamics2}, when $\mathbf{G}=-L_{\mathcal{D}}$,
where $L_{\mathcal{G}}$ is the Laplacian matrix of a directed graph $\mathcal{D}$.
Then, under assumptions (A1)-(A2), the PSD of the input noise, $S_{w}(\omega)$,
can be computed as 
\begin{align*}
S_{w}(\omega) & =\frac{N}{(\mathbf{1}^{T}\mathbf{S}^{-1}\left(\omega\right)\mathbf{1})|h\left(\im\omega\right)|^{2}}.
\end{align*}
\end{corollary} 
\begin{IEEEproof}
This result can be directly obtained from Lemma \ref{input spectrum}
taking into account that the eigenpair $\left(\lambda,\boldsymbol{u}\right)$ for the Laplacian matrix is $\left(0,\boldsymbol{1}\right)$.
\end{IEEEproof}
In general, we can reconstruct the weights of directed edges in a
dynamical network using Algorithm \ref{ExactAlg}.

\begin{algorithm}[t]
\caption{Exact reconstruction of directed networks} \label{ExactAlg}
\begin{algorithmic}[1]
\REQUIRE $h(\im \omega)$, $\mathbf{y}(t)$ from \eqref{dynamics2}, $\widetilde{\mathbf{y}}(t)$ from \eqref{groundedlaplacian}, and any $\omega_{0} \in (-\Omega,\Omega)$;%
\STATE Compute $\mathbf{S}(\omega_{0})$ from $\mathbf{y(t)}$ and ${S}_{w}(\omega_{0})$ using \eqref{input power spectrum};
\FOR{$j=1:N$}%
\STATE Compute $\mathbf{\widetilde{S}}_{j}(\omega_{0})$ from $\widetilde{\mathbf{y}}(t)$;
\FOR{$i=1:j-1$}
\STATE $g_{ji}=\left[S_{w}\left(\omega_{0}\right)\left([\mathbf{S}^{-1}\left(\omega_{0}\right)]_{ii}-[\mathbf{\widetilde{S}}_{j}^{-1}\left(\omega_{0}\right)]_{ii}\right)\right]^{1/2}$;%
\ENDFOR
\FOR{$i=j+1:N$}
\STATE $g_{ji}=\left[S_{w}\left(\omega_{0}\right)\left([\mathbf{S}^{-1}\left(\omega_{0}\right)]_{ii}-[\mathbf{\widetilde{S}}_{j}^{-1}\left(\omega_{0}\right)]_{i-1,i-1}\right)\right]^{1/2}$;%
\ENDFOR
\ENDFOR
\end{algorithmic}
\end{algorithm}

\begin{remark}
It is worth remarking that the reconstruction methods proposed in
the paper do not require the entire power spectra for $\mathbf{S}\left(\omega\right)$
or $S_{w}\left(\omega\right)$, but only the values of these spectral
densities at any frequency $\omega_{0}\in\left(-\Omega,\Omega\right)$.
This dramatically reduce the computational complexity of the reconstruction.
\end{remark}
There are two particular types of networks, namely, undirected and
nonreciprocal networks, in which the computational cost of reconstruction
can be drastically reduced.

\subsection{Exact Reconstruction of Undirected Networks}\label{Undirected Reconstruction}

Consider Problem \ref{Main Problem}, when the connectivity matrix
$\mathbf{G}$ is an unknown (possibly weighted) symmetric matrix.
Then, when an eigenpair $\left(\lambda,\boldsymbol{u}\right)$ is known, we can find the exact structure of the network from the matrix of CPSD's, $\mathbf{S}\left(\omega\right)=\left[S_{y_{i}y_{j}}(\omega)\right]_{1\leq i,j\leq N}$,
and the nodal transfer function, $h\left(\im\omega\right)=c^{T}\left(\im\omega I_{n}-A\right)^{-1}b$,
using the following result:

\begin{theorem} \label{Undirected Network}Consider the networked
dynamical system \eqref{dynamics2}, when $\mathbf{G}=\mathbf{G}^{T}$.
Then, under assumptions (A1)-(A2), we have that 
\begin{equation}
\mathbf{G}=\text{Re}\left\{ h^{-1}\left(\im\omega_{0}\right)\right\} I_{N}+\left(\mathbf{S}^{-1}\left(\omega_{0}\right)S_{w}(\omega_{0})-\text{Im}^{2}\left\{ h^{-1}\left(\im\omega_{0}\right)\right\} I_{N}\right){}^{1/2},\label{undirectedG}
\end{equation}
for any $\omega_{0}\in\left(-\Omega,\Omega\right)$.

\end{theorem}
\begin{IEEEproof}
From Lemma \ref{Power Spectral Densities}, we obtain the following
for $\mathbf{G}^{T}=\mathbf{G}$: 
\begin{align*}
\mathbf{S}^{-1}\left(\omega\right)S_{w}(\omega) & =\frac{I_{N}}{\left|h\left(\im\omega\right)\right|^{2}}+\mathbf{G}^{2}-\frac{\mathbf{G}}{h^{\ast}\left(\im\omega\right)}-\frac{\mathbf{G}}{h\left(\im\omega\right)}\\
 & =\mathbf{G}^{2}-2\text{Re}\{h^{-1}\left(\im\omega\right)\}\mathbf{G}+I_{N}\big(\text{Im}^{2}\{h^{-1}\left(\im\omega\right)\}+\text{Re}^{2}\{h^{-1}\left(\im\omega\right)\}\big)\\
 & =\left(\mathbf{G}-\text{Re}\{h^{-1}\left(\im\omega\right)\}I_{N}\right)^{2}+\text{Im}^{2}\{h^{-1}\left(\im\omega\right)\}I_{N},
\end{align*}
from which we easily derive the statement of our Theorem.
\end{IEEEproof}

Based on Theorem \ref{Undirected Network}, we can reconstruct
the connectivity matrix $\mathbf{G}=\mathbf{G}^{T}$ when we know
an eigenpair of $\mathbf{G}$. The input PSD in \eqref{undirectedG}
can be computed using Lemma \ref{input spectrum}. Notice that this
algorithm does not require grounding the dynamics of the network,
resulting in a reduced computational cost. In particular, the computational
cost is dominated by the computation of $\mathbf{S}\left(\omega_{0}\right)$,
which requires $\mathcal{O}\left(N^{2}L^{2}\right)$ operations, and
its inversion, which requires $\mathcal{O}\left(N^3\right)$, resulting
in a total cost of $\mathcal{O}\left(N^{2}L^{2}+N^3\right)$.

\subsection{Reconstruction of Non-Reciprocal Networks}\label{NonReciprocal Reconstruction}

Another particular network structure that does not require grounding
in the reconstruction method is the so-called nonreciprocal directed
networks. In a \emph{nonreciprocal network}, having an edge
$\left(v_{j},v_{i}\right)\in\mathcal{E}_{d}$ implies that $\left(v_{i},v_{j}\right)\not\in\mathcal{E}_{d}$.
In other words, the connectivity matrix of a purely unidirectional
network satisfies $\text{Tr}(\mathbf{G}^{2})=\sum_{i}\sum_{j}g_{ij}g_{ji}=0$,
since, if $g_{ij}\neq0$, then $g_{ij}=0$ (and assuming there are
no self-loops in the network).

The following Theorem allows the Boolean reconstructing of a nonreciprocal
network. Moreover, if we have access to an eigenpair of $\mathbf{G}$,
this Theorem could be used to perform an exact reconstruction without
grounding the dynamics of the network.

\begin{theorem}\label{Unidirectional Network}

Consider the networked dynamical system \eqref{dynamics2}, with a
connectivity matrix satisfying $\mathbf{G}\geq0$ (nonnegativity)
and $\text{Tr}(\mathbf{G}^{2})=0$ (nonreciprocity). Then, under assumptions
(A1)-(A2), we have that 
\begin{align}
g_{ij}=\max\bigg\{ S_{w}(\omega)\bigg(\frac{[\text{Im}\{\mathbf{S}^{-1}(\omega)\}]_{ij}}{\text{Im}\{h^{-1}(\im\omega)\}}\bigg),0\bigg\},\label{NonreciprocalReconstruction}
\end{align}
for $1\leq i\neq j\leq N$.

\end{theorem}
\begin{IEEEproof}
Under purview of Lemma \ref{Power Spectral Densities}, we obtain
\begin{align*}
\mathbf{S}^{-1}\left(\omega\right)S_{w}(\omega) & =\frac{I_{N}}{\left|h\left(\im\omega\right)\right|^{2}}+\mathbf{G}^{T}\mathbf{G}-\frac{\mathbf{G}}{h^{\ast}\left(\im\omega\right)}-\frac{\mathbf{G}^{T}}{h\left(\im\omega\right)}.
\end{align*}
Taking the imaginary parts, we obtain 
\begin{align*}
\text{Im}\{\mathbf{S}^{-1}\left(\omega\right)S_{w}(\omega)\} & =\text{Im}\{-\frac{\mathbf{G}}{h^{\ast}\left(\im\omega\right)}-\frac{\mathbf{G}^{T}}{h\left(\im\omega\right)}\}=\text{Im}\{h^{-1}\left(\im\omega\right)\}(\mathbf{G}-\mathbf{G}^{T}),
\end{align*}
which entails 
\begin{align*}
\mathbf{G}-\mathbf{G}^{T}=\frac{S_{w}(\omega)}{\text{Im}\{h^{-1}\left(\im\omega\right)\}}\text{Im}\{\mathbf{S}^{-1}\left(\omega\right)\}.
\end{align*}
Given that $\mathbf{G}\geq0$ and the network is nonreciprocal, if
$\left[\mathbf{G}-\mathbf{G}^{T}\right]_{ij}>0$, then $g_{ij}>0$
and $g_{ji}=0$. If $\left[\mathbf{G}-\mathbf{G}^{T}\right]_{ij}<0$,
then $g_{ij}=0$ and $g_{ji}>0$. Finally, if $\left[\mathbf{G}-\mathbf{G}^{T}\right]_{ij}=0$,
then no directed edge between $v_{i}$ and $v_{j}$ exists. These
three conditional statements can be condensed into \eqref{NonreciprocalReconstruction}.
\end{IEEEproof}

Using this Theorem, we can find the the Boolean representation of
$\mathbf{G}$, $\mathbf{B}\left(\mathbf{G}\right)=\left[b_{ij}\right]$,
as follows,
\[
b_{ij}=\begin{cases}
1, & \mbox{if }\frac{[\text{Im}\{\mathbf{S}^{-1}(\omega_{0})\}]_{ij}}{\text{Im}\{h^{-1}(\im\omega_{0})\}}>0,\\
0, & \mbox{otherwise,}
\end{cases}
\]
where $\omega_{0}\in\left(-\Omega,\Omega\right)$. Moreover, if an
eigenvalue eigenvector pair of $\mathbf{G}$ is known, we can recover
$S_{w}\left(\omega_{0}\right)$ using Lemma \ref{input spectrum},
which allows us to recover the value of $g_{ij}$ directly from \ref{NonreciprocalReconstruction}.
Following the analysis of previous algorithms, the computational cost
of the reconstruction of a nonreciprocal directed network is $\mathcal{O}\left(N^{2}L^{2}+N^{3}\right)$.

\section{CONCLUSIONS}

In this paper, we have addressed the problem of identifying the topology
of an unknown directed network of LTI systems stimulated by a wide-sense stationary noise of unknown power spectral density. We have proposed several reconstruction algorithms based on the power spectral properties of the network response to the noise. Our first algorithm reconstructs the Boolean structure of a directed network based on a series of grounded dynamical responses. Our second algorithm recovers the exact structure of the network (including edge weights) when an eigenvalue-eigenvector pair of the connectivity matrix is known. This algorithm is useful, for example, when the connectivity matrix is a Laplacian matrix or the adjacency matrix of a regular graph. Apart from general directed networks, we have also proposed more computationally efficient algorithms for both directed nonreciprocal networks and undirected networks.



\bibliographystyle{IEEEtran}
\bibliography{IEEEabrv,shahin}

\end{document}